# Stoichiometric Epitaxial Strontium Titanate Thin Films on Silicon by High-Temperature Sr Segregation

*Andries Boelen[1,2]†*, Marina Baryshnikova[1]†*, Maxim Korytov[1], Sean R. C. McMitchell[1], Felix Cahyadi[1,3], Christian Haffner[1,4], Clement Merckling[1,2]*

[1] Imec; B-3001 Leuven, Belgium.

[2] Department of Materials Engineering (MTM), KU Leuven; B-3001 Leuven, Belgium.

[3] Physics Department, Jean Monnet University; FR-42100 Saint- Étienne, France.

[4] Department of Information Technology (INTEC), Photonics Research Group, Ghent University; B-9052 Ghent, Belgium.

†Contributing equally.

*Corresponding authors. Email: andries.boelen@imec.be, marina.baryshnikova@imec.be

Keywords: epitaxy, oxides, silicon, stoichiometry recovery.

**Abstract**

Thin-film strontium titanate ($SrTiO_3$, STO) layers grown on silicon require accurate stoichiometry and single-crystalline order to exploit their functional properties optimally. Oxide molecular beam epitaxy can provide an epitaxial interface, but suffers from source oxidation and resulting flux instabilities, yielding only a narrow growth process window for cationic stoichiometry control. Here, we investigate post-growth annealing in oxygen as a pathway to drive the STO layer toward stoichiometry in intentionally Sr-rich epitaxial STO films on silicon (001). Annealing over a broad temperature range revealed two distinct Sr-segregation mechanisms. Below 800 °C, excess Sr segregates toward the surface, forming SrO outgrowths that progressively sublimate at elevated temperatures. Above 800 °C, a second mechanism dominates: Sr accumulates within the interfacial $SiO_2$ layer formed by oxygen diffusion at the STO/Si interface. Together, these mechanisms effectively remove excess Sr from the STO lattice, yielding a more stoichiometric perovskite layer. Our results demonstrate that growing slightly Sr-rich STO templates followed by controlled annealing provides a practical route to improve crystalline quality, offering a scalable strategy for high-quality STO integration on silicon.

## Introduction

Strontium titanate ($SrTiO_3$, STO) thin films grown directly on silicon are attracting increasing interest because they enable functional perovskite oxides to be integrated with mainstream Si technology while retaining wafer-scale compatibility. STO on Si is a versatile epitaxial template for the integration of complex oxides and semiconductors on Si, including $BaTiO_3$ [1,2], $BaBiO_3$ [3], PZT [4], and III–V materials [5]. STO itself is also attractive because of its large permittivity [6,7] and high breakdown strength, making it relevant for applications ranging from energy storage [8] to non-volatile memory devices [9].
In addition, STO recently became a very promising material in cryogenic electro-optic applications [7,10,11]. A ferroelectric state can be accessed through the application of strain [12–14], which enables piezoelectric and electro-optic effects at cryogenic temperatures. Therefore, STO could be used for future devices operating in quantum or low-temperature photonics environments.
For these devices to operate optimally, stoichiometric STO with high crystalline quality is required [10,15]. Solid-source molecular beam epitaxy (MBE) can be considered the most advantageous growth technique for this, thanks to its atomic controllability and uniformity. MBE also provides the possibility to establish an abrupt Si/STO interface because of ultra-high vacuum (UHV) conditions. Although the atomic fluxes can be separately controlled, source oxidation during epitaxy in oxygen environment causes the fluxes to decrease gradually [16–18]. For this reason, cationic stoichiometry control is one of the main challenges in STO epitaxy by MBE. Therefore, it is important not only to improve the growth technique but also to understand the impact of deviation from Sr/Ti stoichiometry on the layer's properties.
In our previous studies, we noticed that Sr- and Ti-rich STO layers have worse structural, electrical, and optical properties than stoichiometric STO [16,19]. Interestingly, they behave differently upon annealing in $O_2$. Ti-rich STO layers degrade as their surface roughness increases along with their crystal structure, which remains poor, while layers with moderate Sr excess seem to even slightly improve after annealing. If an STO layer incorporates a significant amount of extra Sr, annealing leads to serious morphological degradation, including cracks and void formation [16]. The crystallographic properties, however, do improve as the bulk lattice parameter could be restored, as well as a great improvement in crystallinity.
The results of our study on deviations from Sr/Ti stoichiometry in STO layers indicate that Sr can be redistributed in the layer stack during annealing at high temperatures. Similar evidence can also be found in other research works. For example, *Hanzig et al.* studied crystallization dynamics of initially amorphous Sr-rich STO on Si upon annealing in air [20]. It has been shown that at 950 °C, stoichiometric STO with a cubic unit cell could be restored along with a thick $SrSiO_3$ interfacial layer. At the same time, annealing of STO bulk samples in air at 1000 °C leads to the formation of a Sr-depleted surface with SrO islands [21,22]. Furthermore, the diffusion of Sr atoms from bulk STO into the surface of $LaAlO_3$/STO and $NdGaO_3$/STO heterostructures was observed during high-temperature growth (730 °C) [23].
These observations suggest that annealing has a significant influence on the cationic stoichiometry of STO layers and that it can potentially be restored under certain conditions. Therefore, this study aims to examine Sr segregation in epitaxial Sr-rich STO layers on silicon under various annealing conditions and to propose a possible mechanism for this segregation. Finally, understanding the Sr segregation process can open a path towards reproducible high-quality (XRD rocking curve FWHM < 0.1 °, surface roughness $R_a$ < 0.3 nm) stoichiometric STO films on Si with a larger growth process window.

## Experimental methods

Epitaxial STO layers were grown on p-type silicon (001)-oriented substrates using a 200 mm Riber 49 MBE tool equipped with reflection high-energy electron diffraction (RHEED), see Supplementary Figure 1. Sr was evaporated from a dual-filament Knudsen effusion cell, while Ti was generated by an electron beam evaporator. The flux of both metallic molecular beams was calibrated in situ using a quartz crystal microbalance (QCM). For near-stoichiometric STO, the Sr and Ti fluxes were 1.3 Å/s and 0.6 Å/s, respectively. For the Sr-rich layer discussed in this work, the Sr flux was increased to 1.5 Å/s together with a gradual increase of the Sr cell temperature by 15 °C to counteract source oxidation [16]. The Si (001) wafers were cleaned for 90 s in a 2 % HF solution before they were introduced to the UHV growth chamber. Sr-assisted native oxide desorption further prepared the atomically clean surface required for direct epitaxy. Before starting the STO growth, 1/2 monolayer of Sr was first grown, acting as an oxidation barrier between Si and STO. Direct epitaxy of the first 3 nm of STO was performed in molecular oxygen at 350 °C. After this, the growth was paused to switch to atomic oxygen (500 W plasma power) in the growth chamber and to increase the substrate temperature to 550 °C. Under these conditions, the remaining STO epitaxy was completed. The growth time of 50 min resulted in an STO film with a thickness of 75 nm. Finally, a cooling down step to 200 °C was performed over 45 min in oxygen atmosphere, acting as an in-situ annealing treatment to improve oxygen stoichiometry.

Ex-situ postgrowth annealing (PGA) treatments in $O_2$ were carried out using an Annealsys rapid thermal annealing system AS-One. The surface morphology and roughness were investigated using a Bruker Dimension Icon atomic force microscopy (AFM) tool in pulsed force mode. The Sr/Ti ratio in the STO layer was determined by Rutherford backscattering spectrometry (RBS) using a 1.5 MeV He+ ion beam.

For the structural characterization of the STO layers, 2θ–ω scans and reciprocal space mapping (RSM) were performed using a Rigaku SmartLab high-resolution X-ray diffractometer equipped with a high-flux 9 kW rotating-anode X-ray source with Cu as a target material and a HyPix-3000 high-energy-resolution 2D semiconductor detector.

The layer morphology and interfaces were compared for varying annealing conditions by cross-section scanning electron microscopy (XSEM), both a ThermoFisher Verios (3 kV, 0.1 nA, 2.5 mm WD) and a Hitachi SU8000 (10 kV, 10 um, 4.3 mm WD) were used.

Transmission electron microscopy (TEM) was used, together with focused ion beam (FIB) sample preparation, to analyze the microstructure and chemical composition of the films. A Titan$^3$ G2 operated at 200 kV was used for high-angle annular dark-field scanning transmission electron microscopy (HAADF-STEM), annular bright-field STEM (ABF-STEM), dark-field STEM (DF-STEM), and energy-dispersive X-ray spectroscopy (EDS). For *in situ* TEM observation of annealing-induced structural changes in STO, the lamella was mounted on a DENSsolutions Lightning nanochip.

## Results

### Characterization of as-grown Sr-rich STO on Si

A dedicated layer with a large excess of Sr was prepared for better observation of Sr segregation in STO thin films on Si. The applied growth conditions caused a non-uniform distribution of Sr within the layer, with the concentration increasing toward the top surface, as confirmed by secondary ion mass spectrometry (SIMS) analysis (Supplementary Figure 2). Compositional analysis by RBS revealed that the Sr/Ti ratio, averaged over the entire layer, was 1.48 ± 0.04. This implies that the Sr/Ti ratio near the surface was even higher than 1.48. Nevertheless, the surface morphology remained smooth apart from small outgrowths of 1-2 nm height. AFM measurements showed an average roughness of approximately $R_a = 0.2$ nm, which is comparable to that of fully stoichiometric STO layers of similar thickness ($R_a \approx 0.1$ nm) [16] (Supplementary Figure 3). To further investigate the layer morphology and better understand the way of Sr excess incorporation into the STO layer, TEM inspection was performed on the as-grown sample (Figure 1a).

EDS analysis across the TEM lamella confirmed the increase in Sr concentration with layer thickness. Analysis of the STEM and TEM images revealed different morphological regions within the grown layer, each correlated with variations in the Sr/Ti ratio. The initial 5-7 nm of the layer had a near-stoichiometric composition, as evidenced by EDS and a clear alternation of $TiO_2$ and SrO planes in HAADF-STEM images. This indicates a well-ordered perovskite structure.

Since the Sr atomic flux was constantly increased during the layer growth, and given that the solubility of SrO in STO is less than 0.2 mol % [24], the formation of a Sr-rich STO phase was initiated in the system. This phase is commonly referred to as the Ruddlesden-Popper (RP) phase, with a general formula of ($SrO(SrTiO_3)_n$). In the first 10-15 nm region of this non-stoichiometric layer, an increasing Sr excess reaching approximately 25 % was observed. This led to the formation of predominantly vertical defects attributed to the incorporation of additional SrO planes into the structure – RP planar faults [25,26]. The remaining part of the layer exhibited a substantial Sr excess ranging from 25 % to 55 %. Unlike the region below, the extra Sr atoms in this area were most likely incorporated as both vertical and horizontal RP planes. It is interesting to note here that a similar morphology for STO layers with 25 % Sr excess, grown at 650 °C by MBE, was observed in [27]. Furthermore, comparable morphology can also be found in epitaxial Sr-rich STO samples grown by PLD techniques [26,28–30].

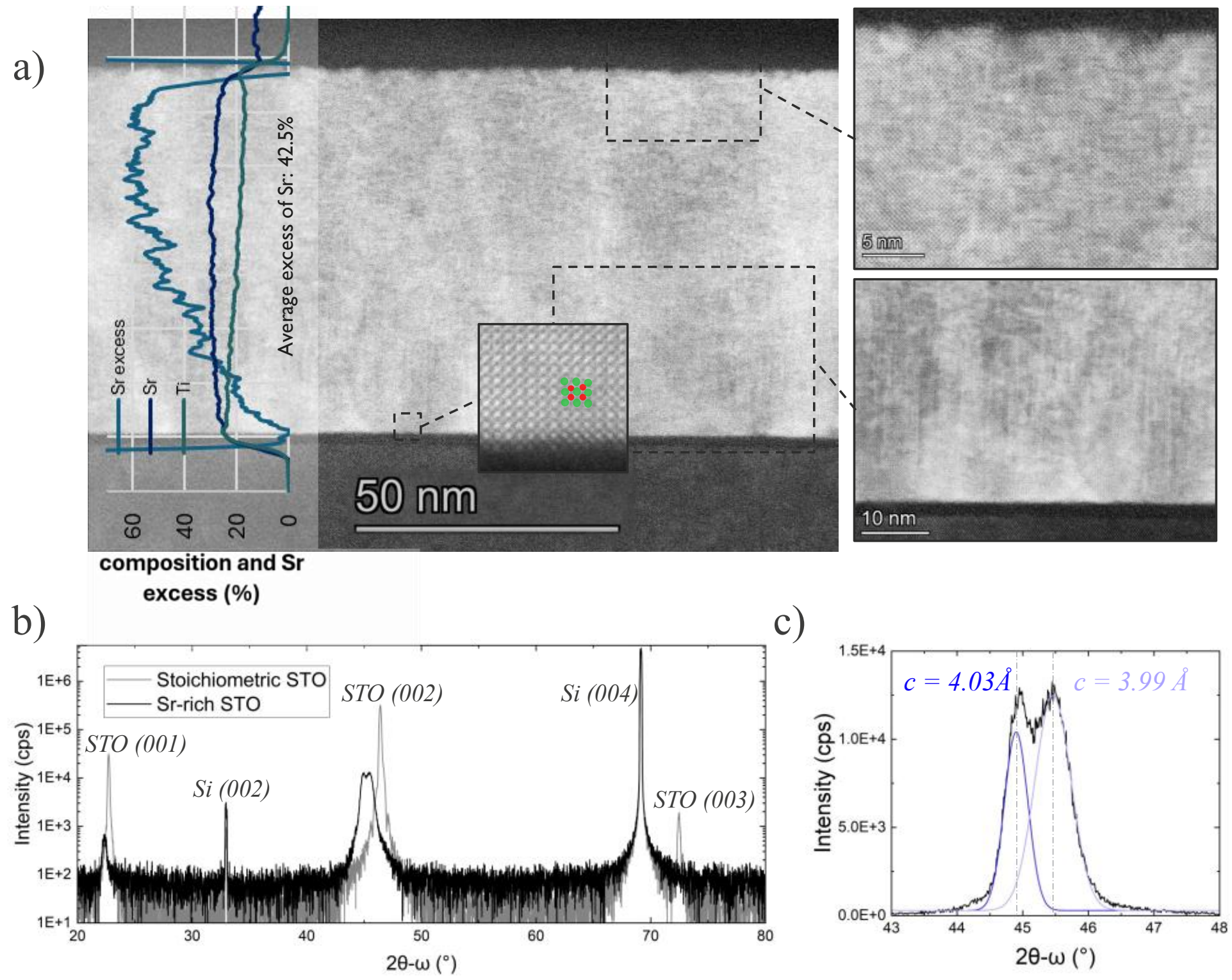

**Figure 1.** Structural characterization of the as-grown Sr-rich STO film. **a)** HAADF-STEM image complemented with EDS The Sr excess within the film increases from the STO/Si interface towards the top surface: from cubic stoichiometric STO layers to vertical RP planar faults to both horizontal and vertical RP planes (see insets). **b)** 2θ-ω XRD scan comparing the Sr-rich film with a stoichiometric STO film of comparable thickness. **c)** Zoomed-in 2θ-ω XRD scan showing two distinct diffraction peaks, corresponding to different lattice parameters within the film.

Despite the significant Sr excess, XRD measurements indicated that the sample could maintain its epitaxial orientation with respect to the Si substrate. No evidence for the formation of polycrystalline STO or other crystalline phases associated with Sr excess, such as Sr or SrO, was found in a standard 2θ-ω scan (Figure 1b). However, the obtained XRD spectrum of the Sr-rich sample exhibited several notable differences compared to that of a fully stoichiometric sample. First, the peak intensity corresponding to crystalline STO was much lower, indicating reduced layer crystallinity. The peak position was shifted to lower 2θ angles, suggesting an increased out-of-plane lattice parameter. This can be attributed to such factors as compressive strain [19], oxygen deficiency [31], and Sr excess [27,28], or their combination.

Furthermore, the (002) diffraction peak consisted of two distinct components and could be deconvoluted into two peaks at 44.90 ° and 45.46 °, corresponding to lattice parameters of 4.03 Å and 3.99 Å, respectively (Figure 1c). The presence of a second peak suggests additional significant changes in the crystalline structure of the Sr-rich STO layer, which could be linked to a tetragonal phase in a strained layer or to the formation of Ruddlesden-Popper (RP) phases [27,32]. In the case of the formation of *ordered* RP phases, multiple higher-order (00l) reflection peaks are expected to be observed in XRD spectra as evidence of an increased degree of structural order [33]. However, since in our case Sr excess was incorporated as randomly positioned RP stacking faults, as evidenced by TEM, these reflections were not detected in our measurements. The second peak could also originate from the formation of out-of-phase boundaries (OPBs). In this case, the separation between the two peaks could be correlated with the density of the OPBs in the layer [34,35]. Interestingly, the double peak disappeared after dipping the sample in diluted BHF, where part of the top STO layer was etched. This indicates the second diffraction peak corresponds to the top STO layers with the most excess Sr (Supplementary Figure 6).

Impact of annealing in oxygen and nitrogen

Annealing Sr-rich STO in an oxygen atmosphere led to a redistribution of Sr within the layer stack, affecting both its crystalline structure and surface morphology. This phenomenon was investigated using RBS, SEM, XRD, and TEM, and the results are summarized in Figure  and Figure 3. Analysis of these findings indicates that Sr segregation in STO layers grown on Si is highly complex, likely involving multiple mechanisms activated under different temperatures and ambient conditions, as discussed below.

At 600 °C, an increase in annealing time has little impact on the layer's crystalline structure, as evidenced by XRD analysis – the peak intensity and position remained almost unchanged unless the annealing time was increased to ≥ 120 min. In this case, a shift of the (002) peak toward higher 2θ angles was observed, accompanied by a small increase in STO peak intensity. At the same time, SEM analysis revealed pronounced morphological changes on the surface. Annealing for 15 min led to the formation of numerous precipitates of triangular shape. Increasing the annealing time to 30 min and above resulted in the formation of outgrowths with more rounded shapes, having 50–70 nm in diameter and around 20 nm in height. The density of these outgrowths depended on the duration of the thermal treatment and was maximal for 60 min of annealing time (Supplementary Figure 4). Dipping of such samples into deionized water (DIW) caused the outgrowths to almost disappear from the surface (Supplementary Figure 6), suggesting they consist primarily of amorphous SrO, which reacts with water to form $Sr(OH)_2$ [22,36,37].

Further increase of the annealing temperature to 700 °C and above induced more notable changes in the layer structure. Particularly, the previously distinct double peak on the XRD pattern began to merge with an increased annealing temperature and/or time, accompanied by an overall increase in peak intensity. At higher temperatures, the peak position further shifted toward 2θ values characteristic of bulk STO, as shown in Figure .

At the same time, annealing at 700 °C for 30 min resulted in the formation of relatively smooth layers, with only a few minor surface outgrowths with much smaller dimensions. When the annealing time was increased to 60 min, the outgrowth density and size significantly increased, making it look similar to the layers annealed at 600 °C. However, a further temperature increase to 800 °C and 850 °C triggered a pronounced transformation in layer morphology. XSEM revealed the formation of a thick interfacial layer between the Si substrate and STO, whose thickness increased with annealing time. Moreover, large, faceted voids were formed in the STO layer, which increased in size with annealing time. Additionally, the surface itself developed multiple pits and cracks after annealing for 60 min at 800 °C and 30 min at 850 °C. This substantial structural and morphological evolution, initiated by a temperature increase to 800 – 850 °C, indicates a high sensitivity of the STO system to thermal processing conditions.

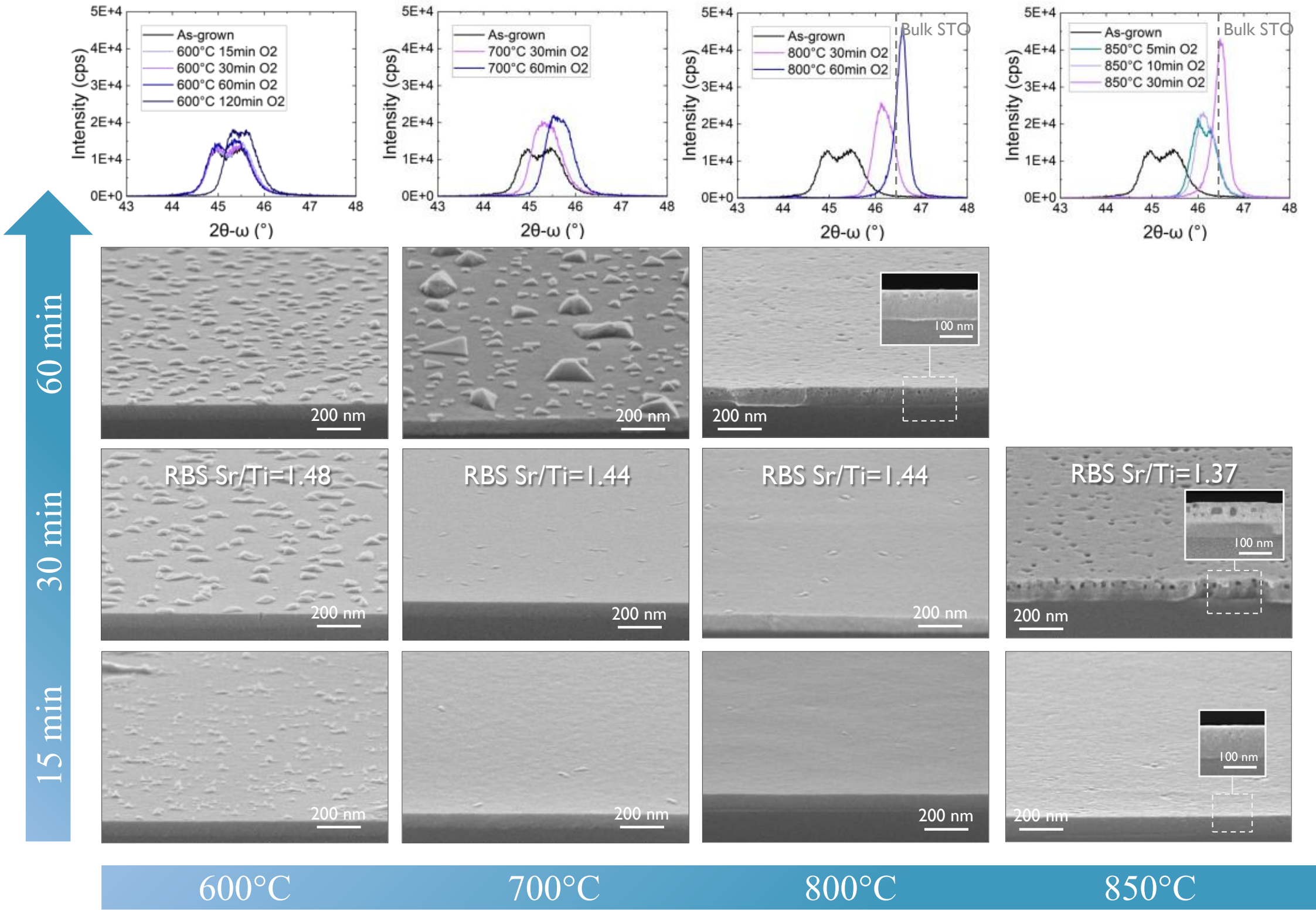

**Figure 2**. SEM and XRD of Sr-rich STO after annealing at varying conditions in $O_2$. At temperatures below 800 °C, large SrO islands are formed on the surface, while only minor structural changes are measured by XRD. At 850 °C (and 800 °C 60 min), no islands but pinholes and voids are gradually formed within the STO layer. Increasing the annealing time at 850 °C to 30 min led to the additional formation of a thick interfacial layer. At the same time, these more aggressive annealing treatments resulted in improved structural properties.

TEM investigation in Figure 3 provided further insights into the structural transformations induced by annealing at 850 °C. An EDS scan performed across the observed 35 nm thick interfacial layer showed that it was composed of silicon oxide with around 20 at. % of Sr. The presence of crystalline phases was not detected in that region. Moreover, the STO layer on top was shown to be completely stoichiometric. High-resolution images of the lattice in that region also confirm the formation of a perfect perovskite structure. The top half of the layer has a large amount of {001}-faceted voids of different sizes, which were also visible in XSEM.

The presence of an interfacial oxide in the STO/Si stack after thermal treatment was also observed in [19,20,38]. However, the exact mechanism of its formation remained rather vague. Our findings suggest that the interfacial layer observed in our samples is predominantly the result of silicon oxidation and possible simultaneous interaction (e.g. interdiffusion and/or chemical reaction) with Sr, as no signal from Ti in the formed oxide layer was detected by EDS. It is also important to note here that the thick oxide layer at the Si/STO interface was only observed for samples having Sr excess and annealed in oxygen ambience at $T \geq 800$ °C. High-temperature annealing in nitrogen or annealing of fully stoichiometric STO samples in oxygen only led to the formation of a thin < 10 nm $SiO_2$ layer at the interface [16].

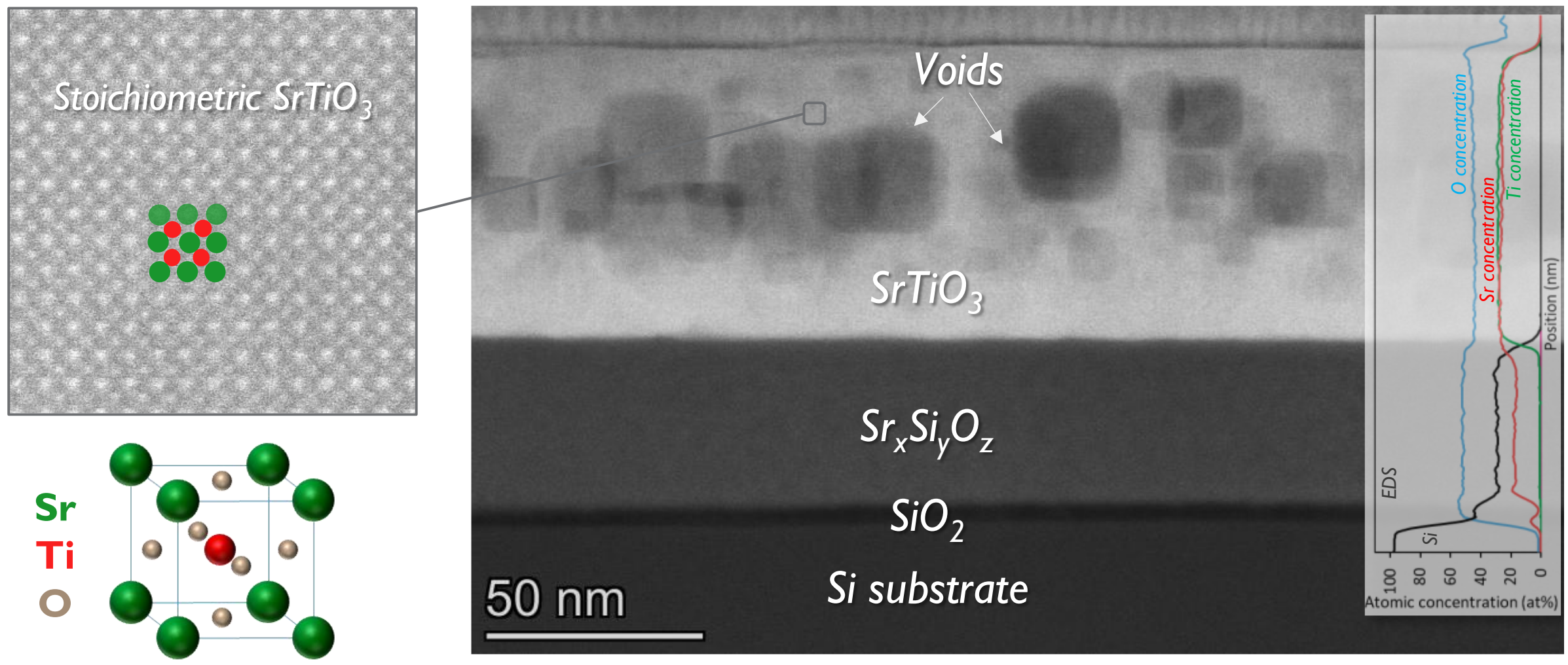

**Figure 3.** TEM of the Sr-rich STO film after annealing at 850 °C in $O_2$ for 30 min, which caused the stoichiometry to be restored in the STO layer. Nevertheless, large voids were formed in the STO layer due to the excessive Sr surplus. Simultaneously, a thick interfacial Sr-rich silicon oxide layer was formed.

Annealing in $N_2$ also resulted in Sr segregation within the layer; however, the effect was less pronounced. SEM investigation indicated that treatments at temperatures up to 800 °C produced visible surface outgrowths, but their density and size were lower than in oxygen-annealed samples. XRD showed possible lattice relaxation as the peak attributed to the (002) reflection was moving to higher angles; however, the characteristic double peak associated with the Sr-rich STO layer persisted (Supplementary Figure 7b). The drastic difference with annealing in $O_2$ was that increasing the temperature to 800 °C in $N_2$ had a detrimental effect on the layer structure. The SEM investigation showed the formation of highly rough layers, and the intensity of the (002) reflection was strongly reduced. At the same time, RBS revealed a decrease in the Sr/Ti ratio to approximately 1.2. This behavior is likely due to decomposition of STO under oxygen-deficient conditions, driven by oxygen loss and simultaneous Sr evaporation (Supplementary Figure 7a).

The results of the structural and morphological investigations discussed above allowed us to propose the presence of two main routes for segregation of excess Sr from STO layers on Si, depending on the annealing temperature in $O_2$. The first one is initiated at relatively mild annealing conditions – 600 °C $\leq$ T < 800 °C - and leads to accumulation of Sr excess near the surface, followed by precipitation in the form of SrO. Increasing the temperature within the indicated range activates diffusion of the excess Sr from the bulk layer as well as sublimation of SrO / Sr from the surface.

The second route for Sr segregation is only observed for annealing in oxygen atmosphere at high annealing temperatures exceeding 800 °C and is caused by the presence of the Si/STO interface. Oxygen plays a special role in this case as the Sr excess is removed from the STO layer through the bottom interface, where it interacts with the forming $SiO_2$ through the formation of a solid solution and/or Sr silicate. The two routes of Sr segregation are discussed in more detail below.

## Discussion

Cation segregation phenomena are well known in perovskite oxides [39–42]. In particular, Sr segregation has been extensively reported for perovskite oxide-based solid electrodes and catalyst materials under high temperatures and oxygen-rich conditions, posing a significant challenge to maintaining their long-term stability. The underlying mechanisms have been thoroughly investigated and summarized in previous studies [43,44]. Sr surface enrichment has also been observed in stoichiometric STO in bulk or thin films when annealed at 1300 °C in UHV [45].

In general, the driving force for the segregation can be attributed to gradients in thermodynamic potentials, which may arise from such factors as strain, defect distribution, oxygen partial pressure, or electric fields. These factors are often dependent on each other, which makes it complex to attribute the observed segregation effects to a single main cause. In our case, the as-grown material exhibits a distorted lattice, suggesting the presence of a certain amount of strain and strain fields. Sr segregation, leading to the formation of a less distorted phase, could therefore contribute to the process. However, the reduced segregation observed during annealing in $N_2$ indicates that strain relaxation alone is unlikely to be the main driver. Comparing the results of annealing in $O_2$ and $N_2$, we conclude that the phenomenon is most likely governed by thermodynamic potential gradients associated with elevated oxygen pressure at the surface.

As it was stated above, based on experimental results, we can distinguish two regimes of Sr segregation depending on temperature. At temperatures below 800 °C, the excess of Sr was diffusing out from the topmost area of the layer with the highest Sr/Ti ratio. Then, after reaching the surface, Sr atoms most likely react with oxygen and form SrO outgrowths, which slowly evaporate into the gas phase. This process can be described by the following reaction: $Sr(s) + \frac{1}{2}O_2(g) \rightleftharpoons SrO\ (g)$. Though there were significant changes in layer morphology, this way of Sr extraction from the initially nonstoichiometric layer can hardly be called efficient, as there was only a minor decrease in Sr/Ti values, (Supplementary Figure 5) comparable to the measurement error. Indeed, results on SrO evaporation published in [46] suggest that SrO sublimation below 800 °C is almost negligible. At the same time, an increase in temperature by only 50 °C leads to an increase in the vaporization rate by an order of magnitude. The more efficient SrO sublimation might explain the reduced density and size of the outgrowths on the surface of the layers annealed at higher than 800 °C compared to the layers annealed at lower temperatures or for shorter times.

Together with the enhancement of SrO sublimation, increasing the annealing temperature to above 800 °C activated diffusion of oxygen into the bulk of the layer [47,48]. At the same time, as evidenced by SEM, faceted voids began to appear only within the 800–850 °C range, and their dimensions increased over time. This observation suggests that the temperature increase favors the initiation of a reaction between Sr and O within the bulk, leading to the formation of volatile SrO species inside the layer volume rather than exclusively at the surface.

A dedicated in-situ TEM and EDS study, where the Sr-rich STO specimen was annealed during inspection, allowed for more insights into the time evolution of the voids and new interfaces in the STO layer at higher temperatures. At the start, a high density of small voids appeared, as shown in Figure 4. With time, their density decreased while individual voids grew larger, suggesting that the smaller voids formed earlier gradually merged as they increased in size.

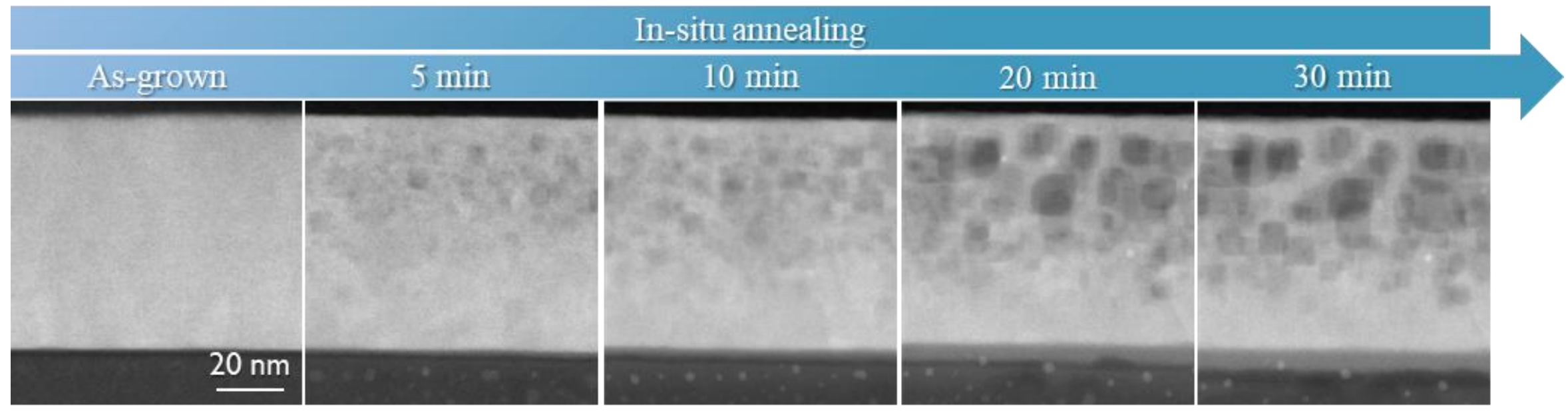

**Figure 4.** Time-dependent in-situ TEM at high temperature. Voids and the interfacial layer are gradually appearing over time because of Sr segregation and Si oxidation, respectively.

Another striking difference for annealing at high temperature was the formation of a thick 35 nm oxide layer at the interface with Si (Figure 3). Since Sr has good solubility in silicon oxide and since at temperatures exceeding 800 °C, the formation of Sr-silicate becomes thermodynamically favorable when SrO and $SiO_2$ coexist [49–53]. EDS analysis indicated that the Sr was uniformly distributed in that layer and its content was reaching 20 at. %, while Si and O were approximately 30 at. % and 55 at. %, respectively. This atomic ratio does not correspond to a well-defined crystalline strontium silicate phase such as $SrSiO_3$ or $Sr_2SiO_4$. Our hypothesis is therefore that the following chemical reactions could take place, leading to the formation of an amorphous Sr-containing silicate [54]:

$$x\ \mathrm{SrO} + y\ \mathrm{SiO_2} \rightleftharpoons \mathrm{Sr_xSi_yO_z}$$

To form these oxide compounds, oxygen influx is required at the interface area. However, as reported in [55,56], the oxygen diffusion in STO can be rather limited. Indeed, for stoichiometric STO annealed at similar conditions, we observed the formation of approximately only 5 nm $SiO_2$ at the interface, which indicates that the effective oxygen concentration at the bottom STO/Si interface is rather low. At the same time, it is surprising to observe the formation of a 35 nm thick oxide for nonstoichiometric STO, indicating a surplus of oxygen. We believe this phenomenon is linked to facilitated oxygen diffusion through the formed voids at temperatures exceeding 800 °C [57,58].

The importance of oxygen in the Sr segregation process can also be concluded from the suppressed formation of the interfacial layer when annealing in situ in TEM. The thickness of this layer was almost three times smaller than for the sample annealed ex-situ at atmospheric pressure. Since the TEM inspection is done under high vacuum conditions, the amount of oxygen available in the sample for interface formation was smaller, leading to reduced Si oxidation at the interface.

The observed Sr segregation phenomenon potentially broadens the process window for achieving high-quality epitaxial STO growth on Si by MBE. Since the standard process lacks a self-regulation mechanism and often deviates from ideal Sr/Ti stoichiometry, we propose to intentionally grow slightly Sr-rich layers. Excess Sr atoms can later be removed through high-temperature annealing in oxygen. It is important, though, that the Sr excess remains within controlled limits, as excessive Sr leads to void formation and layer morphology degradation. To validate this concept, a thin 15 nm STO layer with an Sr excess of approximately 10 % was epitaxially grown on Si. Indeed, after annealing at 850 °C for 30 min, the film restored its Sr/Ti stoichiometry from 1.09 ± 0.02 to 1.00 ± 0.02. Figure 5 summarizes the TEM and AFM results, demonstrating that this strategy enables the formation of a highly crystalline and smooth layer with atomic terraces, thereby opening a pathway toward more reproducible growth of high-quality STO on Si.

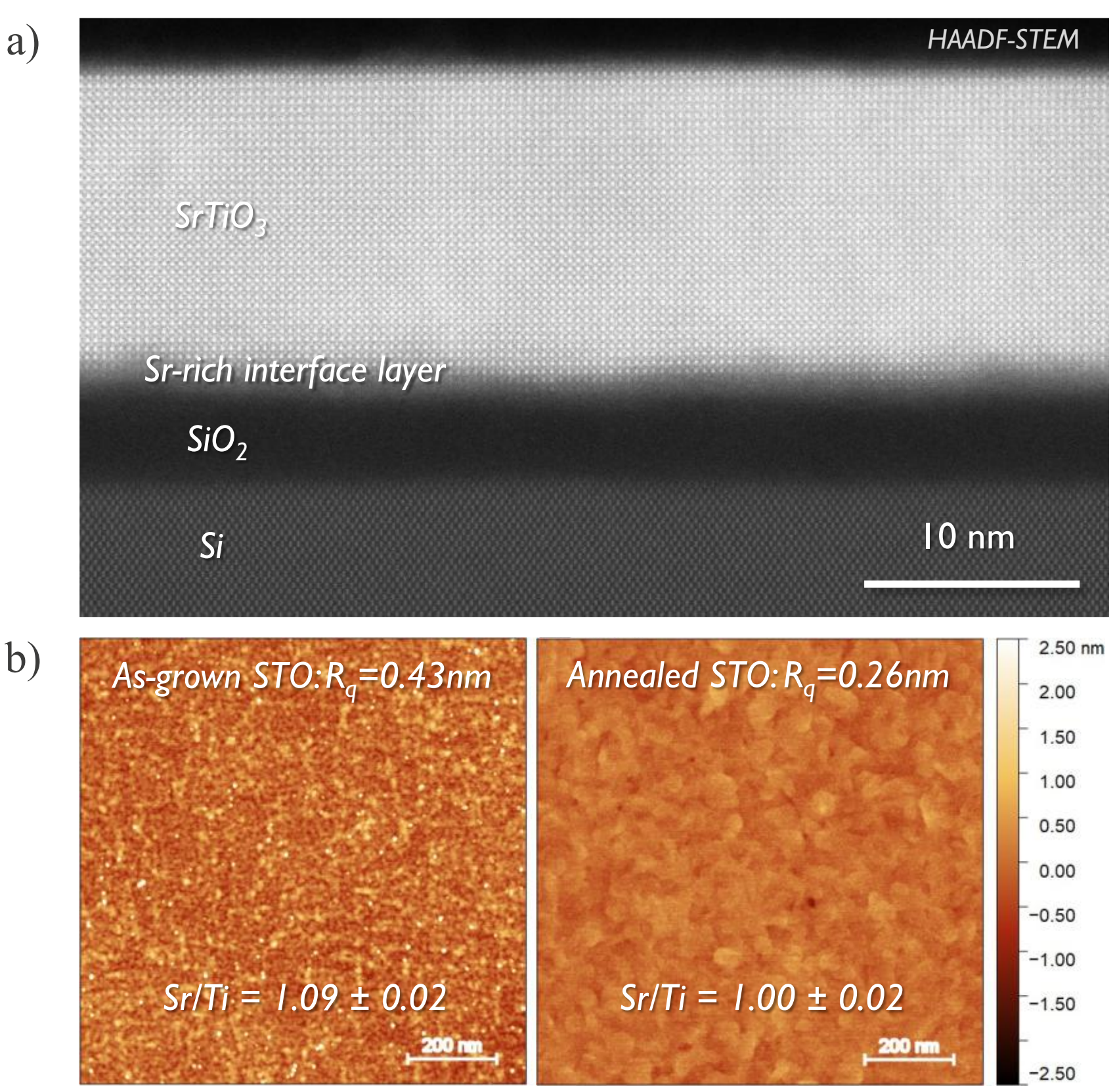

**Figure 5.** High-quality STO templates. **a)** HAADF-STEM of a thin, slightly Sr-rich STO layer (Sr/Ti = 1.09 ± 0.02) after annealing at 850 °C for 30 min (Sr/Ti = 1.00 ± 0.02). Again, a Sr-rich interfacial layer was formed due to Sr segregation, but without any void formation. **b)** AFM after annealing shows a smoothened surface with atomic steps of 1 unit cell high and without any pinholes after annealing.

## Conclusion

Our results indicate that Sr excess in STO can be substantially reduced by annealing the sample at high temperature. The annealing temperature and duration need to be optimized for the Sr excess to ensure Sr segregation while preventing damage to the layer. Introducing oxygen during annealing further enhances the segregation process in two ways: it promotes the formation of SrO, which can subsequently evaporate from the material, and it facilitates Sr segregation at the silicon interface, where annealing drives the removal of excess Sr through the formation of an amorphous Sr-containing silicate.

When the Sr excess is very high, the segregation process can degrade the film morphology, leading to cracks and void formation within the layer. However, if the Sr excess is moderate, annealing can produce a stoichiometric layer with good crystallinity. This approach provides an alternative route for STO growth on Si by MBE, offering a more stable and reproducible process with a broader growth process window.

**Author Contributions**

The manuscript was written through contributions of all authors. All authors have given approval to the final version of the manuscript. Andries Boelen and Marina Baryshnikova contributed equally.

**Funding sources**

This research was funded by the Branco-Weiss Society and the European Research Council (ERC) under the European Union's Horizon 2020 research and innovation program, grant number 864483 NOTICE and grant number 101042414 Q-AMP.

**Acknowledgement**

We thank the MCA department at imec for their support in the AFM, RBS, and SIMS characterization, together with the TEM and XSEM teams.